\documentclass[journal=apchd5,manuscript=letter]{achemso}
\usepackage[version=3]{mhchem} % Formula subscripts using \ce{}

\usepackage{graphicx}
\usepackage{epstopdf}
\usepackage{textcomp}

\author{Sebastian K. H. Andersen}
\author{Anders Pors}
\author{Sergey I. Bozhevolnyi}
\email{seib@iti.sdu.dk}
\affiliation[University of Southern Denmark]
{Department of Technology and Innovation, University of Southern Denmark, Niels Bohrs All{\'e} 1, DK-5230 Odense M, Denmark}

\title[Gold Photoluminescence Engineering]
  {Gold Photoluminescence Wavelength and Polarization Engineering}

\abbreviations{GSPR, gap surface plasmon resonance; GSP, gap surface plasmon; IR, infrared; LSP, localized surface plasmon; PL, Photoluminescence; SEM, scanning electron microscopy; SI, supporting information; Ti, titanium.}


\begin{document}

%\begin{tocentry}

%Some journals require a graphical entry for the Table of Contents.
%This should be laid out ``print ready'' so that the sizing of the

%Inside the \texttt{tocentry} environment, the font used is Helvetica
%8\,pt, as required by \emph{Journal of the American Chemical

%The surrounding frame is 9\,cm by 3.5\,cm, which is the maximum
%permitted for  \emph{Journal of the American Chemical Society}
%graphical table of content entries. The box will not resize if the
%content is too big: instead it will overflow the edge of the box.

%This box and the associated title will always be printed on a
%separate page at the end of the document.

%\end{tocentry}

\begin{abstract} 
We demonstrate engineering of the spectral content and polarization of photoluminescence (PL) from arrayed gold nanoparticles atop a subwavelength-thin dielectric spacer and optically-thick gold film, a configuration that supports gap-surface plasmon resonances (GSPRs). Choice of shapes and dimensions of gold nanoparticles influences the GSPR wavelength and polarization characteristics, thereby allowing us to enhance and spectrally mold the plasmon-assisted PL while simultaneously controlling its polarization. In order to understand the underlying physics behind the plasmon-enhanced PL, we develop a simple model that faithfully reproduces all features observed in our experiments showing also good quantitative agreement for the PL enhancement.
\end{abstract}

\begin{tocentry}

	\centering
		\includegraphics{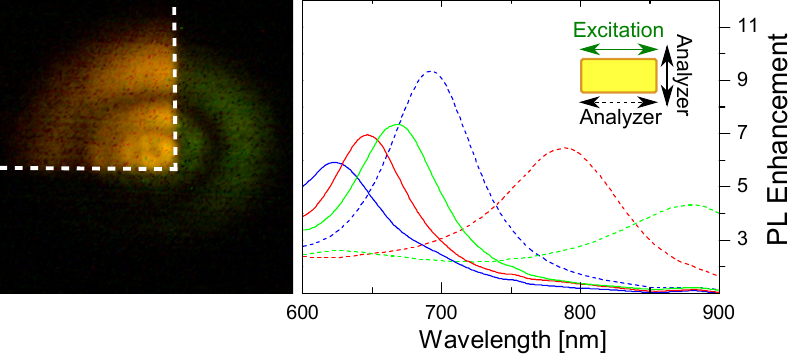}

\end{tocentry}

%%%%%%%%%%%%%%%%%%%%%%%%%%%%%%%%%%%%%%%%%%%%%%%%%%%%%%%%%%%%%%%%%%%%%\redline{Mooradin}

Photoluminescence (PL) from gold was first brought to attention in 1969 by 
 Mooradian\cite{Mooradin1969}, who observed broadband unpolarized emission from a plane gold film, which was also independent on the pump light polarization. 
PL from gold has conventionally been attributed to a three-step process consisting of photoexcitation of electron-hole pairs by excitation of d-band valence electrons to the s-p conduction band, relaxation of electron-hole pairs, and finally photon emission by electron-hole recombination\cite{Apell1988}.
 The radiative recombination of electrons and holes from a plane gold film was reported to be an extremely inefficient process with an estimated quantum yield of $\eta_0\sim10^{-10}$.
Experimental observations of PL enhancement from rough gold surfaces was later theoretically treated by Boyd \textit{et al.} \cite{SurfaceRoughness}.
 By approximating surface roughness as a random collection of noninteracting hermispheriods, PL enhancement was assigned to the excitation of localized surface plasmons (LSPs), which resulted in improvement of both excitation and emission efficiency by enhancing the local electric field at the respective wavelengths. 
  The electric field enhancement picture, presented by Boyd \textit{et al.}, has been widely adopted for the description of PL from nanoparticles for both linear \cite{Chatdanai2014,Jaebeom2004} and nonlinear excitation \cite{Bouhelier2003,Beermann2008}.
  Experimental evidence of the plasmon-enhanced PL has been demonstrated with single gold nanoparticles, for which the PL spectra were strongly modulated by a pronounced peak at the wavelength of the LSP \cite{Bouhelier2005,Micael2003}.
  Very large enhancements in PL quantum yield (up to $10^6$) have been reported when conducting ensemble measurements of gold nanorods in solutions\cite{Mohamed2000}, and tuning the transverse and longitudinal LSPs to the excitation and emission wavelength, respectively. 
   
\begin{figure}
	\centering
		\includegraphics{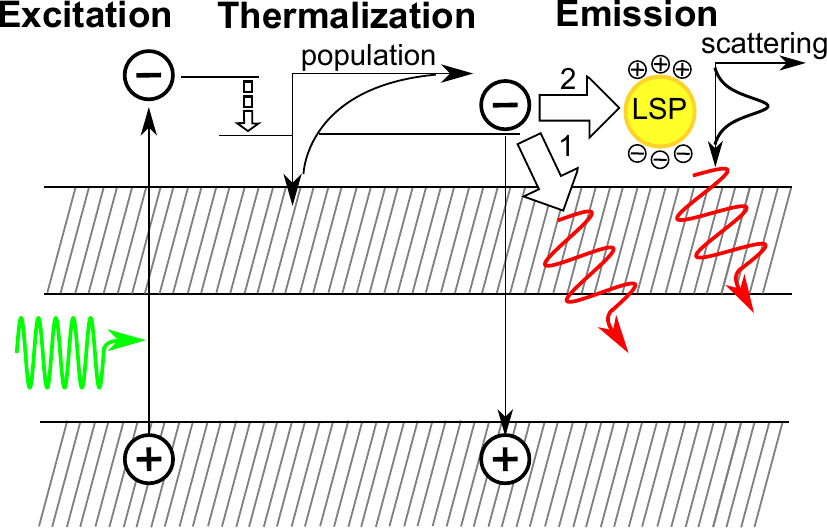}
		\caption{Sketch of the processes involved in photoluminescence from gold nanostructures.
		}
	\label{fig:Figure5}
\end{figure}

More recently, increased attention has been given to the mechanism of radiative recombination of electrons and holes.
The radiative decay of electron-hole pairs has been suggested\cite{Dulkeith2004} to consist of a direct contribution from recombination into photons and an indirect contribution from excitation of an LSP that subsequently decays via both absorption (ohmic loss) and radiation in the form of PL (Figure \ref{fig:Figure5}). 
The mechanism of plasmon-enhanced PL as being associated with the LSP excitation is largely confirmed, since many of the LSP properties were found to be imprinted into the corresponding PL characteristics. 
For example, LSP wavelengths seen in scattering were found to practically coincide with the peaks in PL spectra\cite{Shahbazyan2013,Walsh2013,Mathias2008,Ying2012}. 
Also, the temporal delay of plasmon-enhanced PL emission with respect to its excitation is $\leq{50}$\,fs\cite{Oleg2003} and, thus, of the same order as the LSP lifetime\cite{Sonnichsen2002}.
 Finally, it has recently been demonstrated that plasmon-enhanced PL is polarized along the resonant axis of gold nanorods\cite{Alexei2011,Mustafa2012} or transverse to a silver nanowire on a dielectric spacer/gold surface\cite{HailongHu2013}.
  In this regard, it is worth noting that polarization resolved PL studies of nanorods have been limited to the long-axis LSP excitation, although utilization of the transverse LSP may improve the overall control of  PL polarization.

In this work, we demonstrate the PL wavelength and polarization engineering by exploiting arrayed gold nanoparticles atop a subwavelength-thin dielectric spacer and optically-thick gold film, a configuration that supports gap-surface plasmon (GSP) resonances\cite{Michael2012}. Spectral positions of GSP resonances (GSPRs) can be straightforwardly controlled by choosing shapes and dimensions of the nanoparticles\cite{pors2013_1,roberts2014}. Moreover, GSP-based structures in this geometry are characterized not only by large electric field enhancements at resonance, but also by ease of nanofabrication with one step of electron-beam lithography being involved\cite{Michael2012,roberts2014}. These properties make GSP-resonator arrays attractive for plasmon-enhanced PL wavelength and polarization engineering. Especially, the usage of arrayed rectangular shaped nanoparticles (i.e. nanobricks) allows us to illustrate the excitation dependent interplay of two non-degenerate LSPs (associated with the resonant excitation of two orthogonally propagating GSPs) and its use for the gold PL engineering. In order to understand the underlying physics responsible for spectral and polarization properties of the plasmon-enhanced PL, we develop an intuitive model incorporating the local field enhancement effects at the excitation wavelength and explicitly relating the PL occurrence to the excitation of an LSP(Figure \ref{fig:Figure5}), whose radiative decay generates the enhanced PL with the corresponding spectral and polarization properties. This model faithfully reproduces all PL features observed in our experiments, including the PL dependence on the excitation polarization that has recently been reported\cite{HailongHu2013,HailongHu2012}. Moreover, it is also found in good quantitative agreement when evaluating the PL enhancement, thus providing design guidelines for the PL wavelength and polarization engineering.   

\section{Results and discussion}
\begin{figure}
	\centering
		\includegraphics{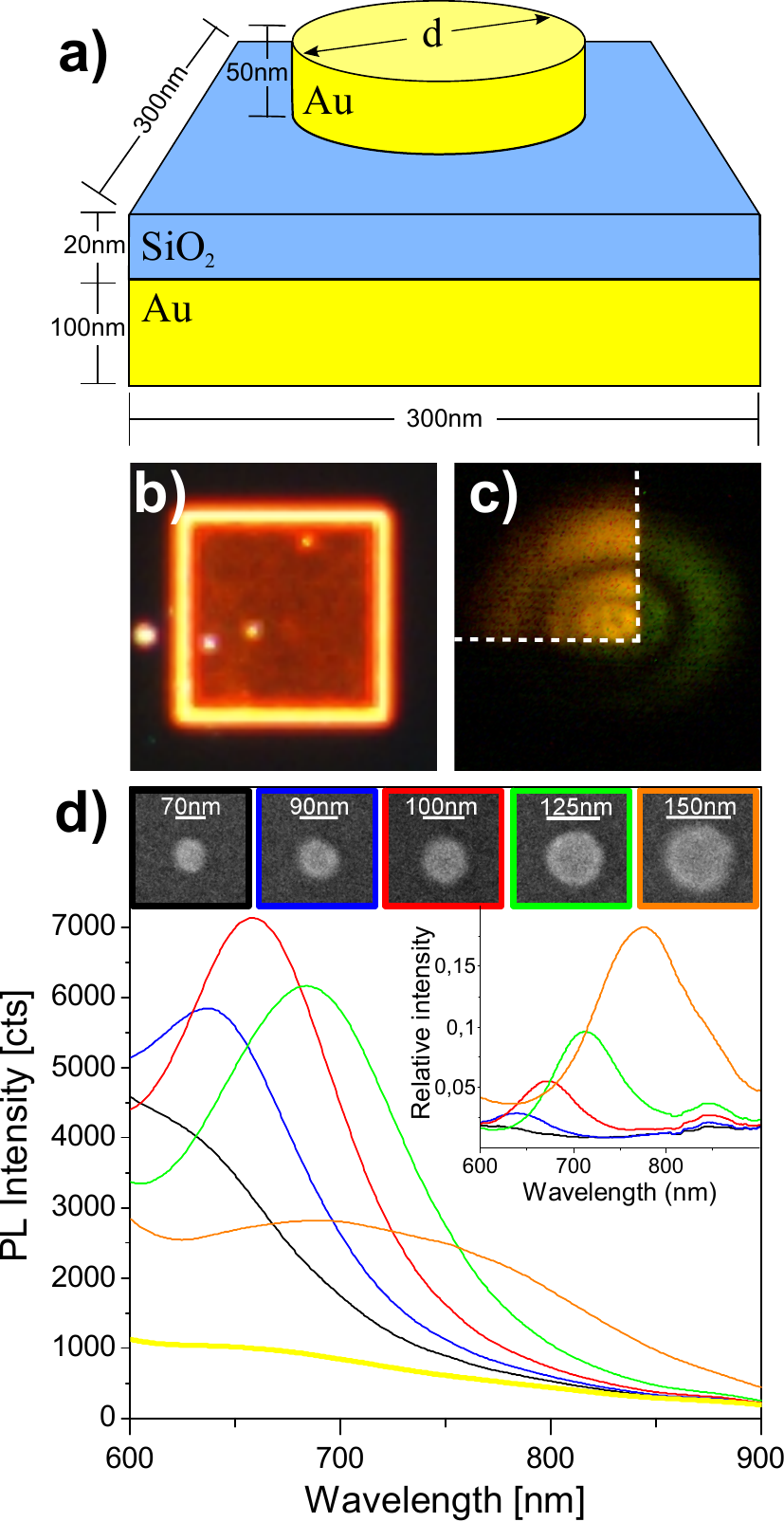}
		\caption{(a) Schematic of unit cell dimensions for nanodisk GSP-resonator array. (b) Dark-field image of $50\times50$\,\textmu m$^2$ nanodisk array with $d=100$\,nm. (c) Plasmon-enhanced PL arising from the overlap of the excitation laser spot with the nanodisk array imaged in (b)(indicated by the dashed boundary), and the surrounding plane SiO$_2$/gold surface. (d) PL spectra of five nanodisk arrays and plane SiO$_2$/gold surface (thick yellow line); the corresponding scattering spectra relative to silver mirror reflection is given in inset, with SEM micrographs of the nanodisks given in top panel.		 
		}
	\label{fig:Figure1}
\end{figure}

\paragraph{Experimental Results}
Let us start by considering plasmon-enhanced gold PL arising from a GSP-resonator array of circular shaped nanoparticles (i.e. nanodisks) illuminated at a wavelength of 532\,nm. The GSP-resonator configuration consists of a 100-nm-thick gold film on a silicon substrate with a 20-nm-thin SiO$_2$ dielectric spacer layer and a 50\,nm top gold layer of nanodisks, defined by electron beam lithography, in a 50$\times$50\,\textmu m$^2$ array with a 300\,nm period (Figure \ref{fig:Figure1}a).
The structure features Fabry-Perot like resonances\cite{Bozhevolnyi2007}, originating from propagating GSP modes that for certain wavelengths (or disk sizes) form standing wave LSPs due to efficient reflection at disk boundaries. Here, we only consider the fundamental GSPR, which is (within the metamaterial community) also known as the magnetic resonance\cite{yuan2007}, whose LSP wavelength can be simply controlled by varying the diameter of the disks. In fact, we vary the LSP wavelength from $\sim{600}-780$\,nm by varying the average nanodisk diameter from $\sim70-150$\,nm, as evident from correlating scanning electron microscopy (SEM) images and dark-field scattering spectra (Figure \ref{fig:Figure1}d). 
Note also a distinct dark-red color of a nanodisk array, observed with a dark field microscopy, originating from the strong scattering of light at the corresponding LSP wavelength (Figure \ref{fig:Figure1}b). 
The LSP modes have apparent influence on the strength and spectral distribution of the PL signal, as illustrated by imaging the PL with the excitation laserspot only partially overlapping the nanodisk array (Figure \ref{fig:Figure1}c). 
In order to quantify the influence of GSP-resonator arrays on the PL signal, we have recorded the PL spectrum for each array (Figure \ref{fig:Figure1}d) using a linear polarized pulsed laser at wavelength of 532\,nm, weakly focused on the sample to ensure a broad excitation covering the center portion of the array.
The PL count was found linearly dependent on the excitation power, confirming that the observed PL is related to single photon absorption.
The PL from the plane gold film is seen to gradually decrease in intensity towards long wavelengths, since the population of excited charge carriers decreases for energies being further away from the excitation. Contrary, the PL spectra from the nanodisk arrays are strongly modulated, featuring higher photon counts and  
reaching maxima that are progressively red shifted when increasing the disk diameter.
The plasmonic origin of the observed PL enhancement is confirmed by a direct comparison of the PL and scattering spectra (Figure \ref{fig:Figure1}d), that reveal a close resemblance between spectral peak positions.

\begin{figure}
	\centering
		\includegraphics{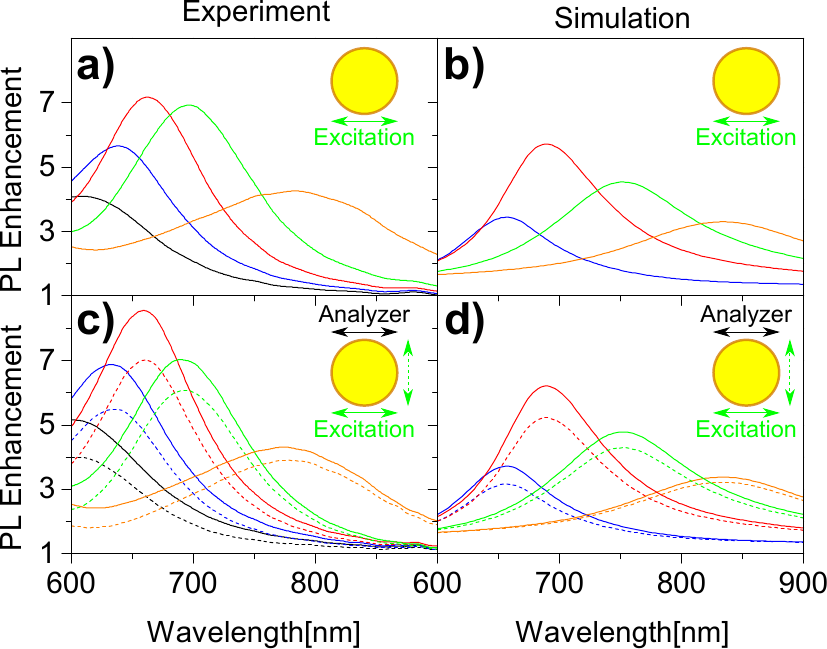}
		\caption{PL enhancement of nanodisk GSP-resonator arrays relative to plane SiO$_2$/gold film, with array color identifier established in Figure \ref{fig:Figure1}d, without analyzer (a) experiment and (b) simulation, and with analyzer for excitation polarized parallel (solid) and perpendicular (dashed) to analyzer given for (c) experiment and (d) simulation. In simulations, disks diameters are 80\,nm (blue), 100\,nm (red), 120\,nm (green), and 140\,nm (orange). }

	\label{fig:Figure2}
\end{figure}

The decreasing PL intensity for LSP wavelengths below 650\,nm, we attribute to increased non-radiative damping of the LSP from interband transitions in the gold, which is well known to set in for wavelengths $<{650}$\,nm\cite{Johnson1972}.  Interestingly, the PL intensity also falls off at a significant rate for the LSP wavelengths red-shifted beyond 650\,nm, despite the fact that the radiative damping of the LSP is increasingly dominating over the non-radiative decay channel with increasing particle size. 
The same effect has previously been attributed to the decreasing population of non-equilibrium charge carriers for plasmon excitation at long wavelengths\cite{HailongHu2012} . In order to account for charge carrier population effects, we obtain the PL enhancement of the respective nanodisk arrays (Figure \ref{fig:Figure2}a) by normalizing the PL spectra with an identically measured PL spectrum of the plane gold film, for which the PL intensity should be directly proportional to the excited charge carrier population.  It is important to note, that the PL spectra arise from a large excitation spot area on the array containing both areas of weak PL (gold film) and strong PL (nanodisks). The PL enhancement therefore represents the surface average enhancement, which is very different from single nanoparticle PL enhancement for which only the nanoparticle area is considered. The PL enhancement is decreasing for LSP wavelengths  $>$\,650nm, hereby indicating a secondary effect (beyond a decreasing charge carrier population) is contributing to the decline in PL intensity.
As will be clear from the proposed theoretical model, plasmon-enhanced PL results from the spatial overlap between the absorption profile within the metal, at the excitation wavelength, and LSP mode profile. The decline in PL enhancement at long LSP wavelengths is a consequence of decreasing overlap, resulting from the gold becoming more metallic at long wavelengths, thus lowering the fraction of LSP mode energy in the metal. Furthermore, the enhanced absorption via off-resonant LSP excitation at the pump wavelength is progressively diminished with increased LSP red-shift from excitation wavelength. Finally, we note that the PL peaks, in accordance with the LSP resonances observed in scattering, spectrally broaden with increasing particle size. The broadening of the line width results from increasing LSP damping, for which especially radiative damping intensify with particle size. As such, one may narrow the spectral width of PL peaks by suppressing the influence of the electric dipole moment of the LSP mode\cite{Pors2010}. The cost, however, is a reduced PL enhancement because of the decreased LSP quantum yield. 

We now consider the polarization-dependent properties of the nanodisk arrays. By inserting an analyzer in the PL collection path, the PL polarization was checked as a function of excitation polarization (Figure \ref{fig:Figure2}c). Despite the fact that nanodisks possess two orthogonal degenerate LSP modes, the PL is slightly polarized along the direction of excitation, as PL intensity is consistently strongest in the parallel analyzer-excitation configuration. We would like to emphasize that the same conclusion can be reached by utilizing an orthogonally oriented analyzer [see Figure S1 in Supporting Information (SI)]. The observation can be explained by the absorption profile in the metal spatially overlapping more strongly with the co-polarized LSP mode profile compared to the orthogonally polarized, thus resulting in an uneven excitation of the LSP modes.
As a measure of the polarization contrast of PL, we define the quantity $\alpha=|I_{\parallel}-I_{\perp}|/|I_{\parallel}+I_{\perp}|$ at the LSP wavelength, where $I_{\parallel}$ and $I_{\perp}$ are, respectively, the PL intensity polarized parallel and perpendicular to excitation. For the nanodisk arrays, we find a value of $\alpha\simeq 0.1$ for the peak PL intensity (Figure \ref{fig:Figure2}c; red curve), with the quantity decreasing for longer wavelengths.

\begin{figure}
	\centering
		\includegraphics{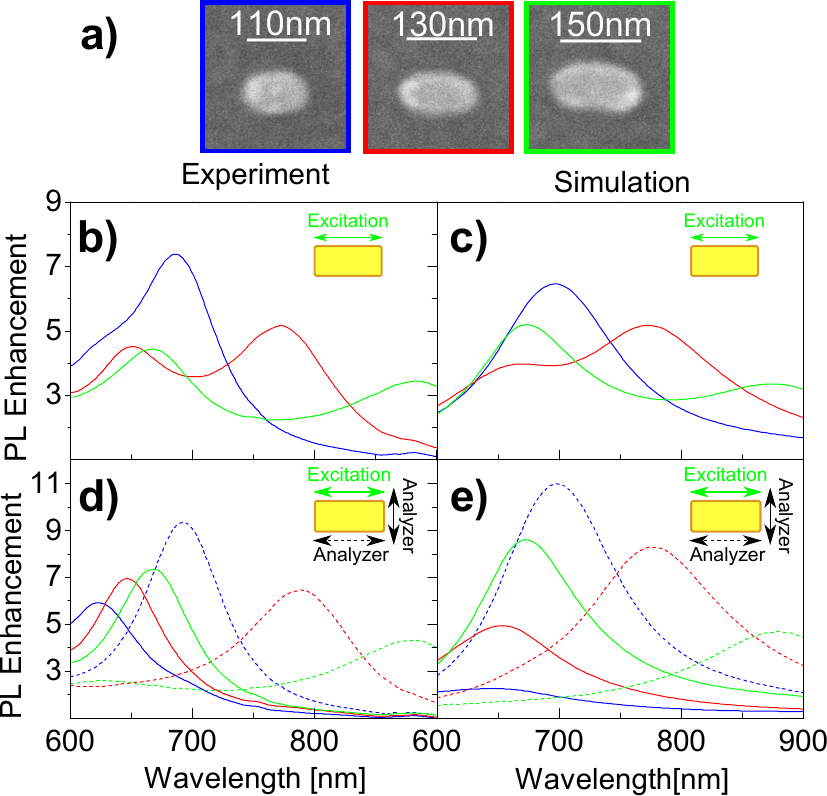}
		\caption{ (a) SEM micrographs of nanobrick GSP-resonators with frame color identifier blue, red, and green. (b-e) PL enhancement of nanobrick GSP-resonator arrays relative to plane film for excitation polarized along the long axis of the brick; (b) measured and (c) calculated PL spectra for detection without analyzer; (d) experiment and (e) simulation for PL spectra detected with analyzer along the long axis of the brick (solid) and for orthogonal orientation (dashed). In simulations, the geometrical cross sections of nanobricks are $80\times40$\,nm$^2$ (blue), $105\times55$\,nm$^2$ (red), and $130\times70$\,nm$^2$ (green).}
	\label{fig:Figure3}
\end{figure}

\begin{figure}
	\centering
		\includegraphics{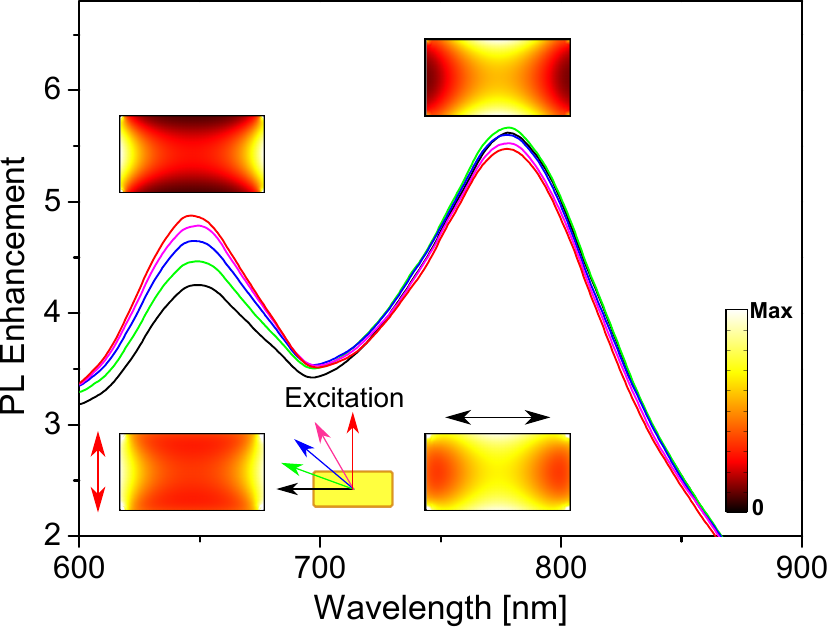}
		\caption{Measured PL enhancement spectra of nanobrick GSP-resonator array for the excitation polarization angled with respect to the long axis of the brick at 0$^\circ$ (black), 20$^\circ$ (green), 40$^\circ$ (blue), 60$^\circ$ (pink) and 90$^\circ$ (red), as illustrated by colored arrows in the lower part of the figure. Insets show simulations of the field intensity distribution ($|E|^2$) in the gold nanobrick at the LSP resonances (upper panels) and at the excitation wavelength for polarization along the two main axes (lower panels). 
		}
	\label{fig:Figure4}
\end{figure}

It is clear from the above discussion that nanodisk arrays allow for an enhanced and spectrally-engineered PL emission, with the polarization contrast being weak due to the degeneracy of orthogonal LSP modes. As a way to control the polarization properties of plasmon-enhanced PL, we now study nanobrick GSP-resonator arrays, featuring two orthogonal non-degenerate LSPs that can be spectrally tuned by adjusting the size and aspect ratio of the nanobricks. 
The nanobricks were prepared with the same layer thickness and array configuration as the nanodisk arrays. Three arrays with average geometrical cross sections of nanobricks being $110\times 70$\,nm$^2$, $130\times 72$\,nm$^2$, and $150\times 77$\,nm$^2$ (Figure \ref{fig:Figure3}a) were investigated  (Figures \ref{fig:Figure3}b and \ref{fig:Figure3}d). The corresponding PL spectra (Figure \ref{fig:Figure3}b) show two peaks resulting from the two orthogonally polarized LSP modes associated with the short and long axis, as confirmed by scattering measurements (see Figure S2 in SI). Consequently, the polarization resolved PL spectra feature individual peaks related to the correspondent LSPs, resulting thereby in strongly polarized emission at both LSP wavelengths (Figure \ref{fig:Figure3}d). For excitation polarized along the long axis of the nanobrick, the polarization contrast at the corresponding long wavelength LSP is as large as $\alpha\simeq0.7$ (Figure \ref{fig:Figure3}d; red curves), with an average value of $\sim 0.6$ for the three cases considered. For the short wavelength LSP, the polarization contrast is slightly lower with $\alpha\simeq 0.5$ for the two largest nanobrick configurations, while the latter configuration (Figure \ref{fig:Figure3}d; blue curves) demonstrates $\alpha\simeq 0.3$. It is evident from the presented $\alpha$-values that nanobrick arrays facilitate the possibility (within a desired wavelength range) to engineer the polarization of PL from being completely unpolarized to strongly polarized by choosing nanobricks of proper size and aspect ratio. In this regard, it is worth noting that the nature of PL does not allow for complete polarization contrast ($\alpha=1$) since the omni-present PL contribution from the direct radiative decay of an electron-hole pair, although it might be small compared to the plasmonic contribution, is unpolarized.
As a final comment, it should be emphasized that the total PL spectrum of nanobrick arrays can be modified by varying the angle of the excitation polarization relative to the nanobrick axes (Figure \ref{fig:Figure4}), which alters the relative contribution from the two non-degenerate orthogonal LSPs. 

Similar to the case of the nanodisk array, the largest contribution to PL from an LSP mode is achieved when the excitation polarization is aligned with the associated axis, as the absorption profile in the metal achieves the largest spatial overlap with the LSP mode profile (see insets in Figure \ref{fig:Figure4}). The PL dependency on excitation polarization is observed to weaken when the LSP wavelength moves away from excitation wavelength. Note that this effect has also been observed for the other two nanobrick arrays (see Figure S3 in SI). 

\paragraph{Theoretical discussion}
As a way to better understand the underlying physics behind PL spectra from gold nanostructures, we develop a theoretical model that (despite its simplicity) is able to account for all the features observed in the above experiments. Following the discussion of related work\cite{Dulkeith2004,Walsh2013,HailongHu2012}, the generation of PL is considered as a three-step process (Figure \ref{fig:Figure5}) involving excitation of electron-hole pairs, thermalization (giving rise to broad distribution of electron energies), and relaxation either via direct radiative recombination or indirectly by excitation of an LSP that may subsequently decay radiatively as PL or nonradiatively (ohmic loss). As a starting point, we assume the three processes to be independent, meaning that for a plane gold film, in which the plasmonic relaxation path is not an option, we may write the position-dependent PL intensity as 
\begin{equation}
I_{\textrm{PL}}^{\textrm{film}}(\omega_\textrm{ex},\omega,\mathbf{r})=U(\omega_\textrm{ex},\mathbf{r})f(\omega_\textrm{ex},\omega)\eta_0,
\label{eq:PLfilm}
\end{equation} 
where $\omega$ is the angular frequency, $\omega_\textrm{ex}$ is the excitation frequency, $\mathbf{r}$ is the coordinate vector, $U=\tfrac{1}{2}\omega_\textrm{ex}\varepsilon_0\varepsilon_\textrm{au}^{''}(\omega_\textrm{ex})|\mathbf{E}(\mathbf{\omega_{ex},r})|^2$ is the Ohmic heating density that represents the rate of excitation, $\varepsilon_0=8.854\cdot10^{-12}$\,F/m is the vacuum permittivity, $\varepsilon_\textrm{au}^{''}$ is the imaginary part of the relative gold permittivity, $\mathbf{E(\omega_{ex},r)}$ is the local electric field, $f$ represents the probability density distribution of excited electron-hole pairs, and $\eta_0\sim10^{-10}$ is the quantum yield. Note that $f$ has the property that $\int_0^{\omega_\textrm{ex}}f(\omega_\textrm{ex},\omega)\mathrm{d}\omega=1$, which ensures that the total PL intensity (obtained by integrating equation \ref{eq:PLfilm} over volume and frequency) divided by the total power dissipated in the metal at the excitation frequency equals $\eta_0$, as originally defined by Mooradin\cite{Mooradin1969}. If we now consider a metallic nanostructure, there is a probability $P(\omega,\mathbf{r})$ for relaxation of excited electron-hole pairs by excitation of an LSP, which leads to the position-dependent PL intensity
\begin{align}
I_{\textrm{PL}}^{\textrm{ns}}(\omega_\textrm{ex},\omega,\mathbf{r})&=U(\omega_\textrm{ex},\mathbf{r})f(\omega_\textrm{ex},\omega)\left[\eta_0\left(1-P(\omega,\mathbf{r})\right)+P(\omega,\mathbf{r})\eta_{\textrm{lsp}}\right], \nonumber \\
& \approx U(\omega_\textrm{ex},\mathbf{r})f(\omega_\textrm{ex},\omega)\left[\eta_0+P(\omega,\mathbf{r})\eta_{\textrm{lsp}}\right],
\label{eq:PLns}
\end{align}
where the validity of the last equality owes to $\eta_{\textrm{lsp}}\gg \eta_0$, since the quantum yield $\eta_{\textrm{lsp}}$ of the LSP is determined by the ratio of scattering to extinction cross section at resonance and, hence, typically $\eta_{\textrm{lsp}}> 0.1$ for particles larger than a few tenths of nanometers\cite{Dulkeith2004}. In order to proceed, we note that the probability of an excited electron-hole pair at position $\mathbf{r}$ to relax via LSP excitation is proportional to the local density of states of the LSP mode at that position\cite{Viarbitskaya2013}. Here, we approximate the probability of LSP excitation by the local field intensity enhancement at resonance, i.e., 
\begin{equation}
P(\omega,\mathbf{r})=\beta\frac{|\mathbf{E}(\omega_{\textrm{lsp}},\mathbf{r})|^2}{E_0^2}N(\omega),
\label{eq:p}
\end{equation}
where $\beta$ is a proportionality factor and $\mathbf{E}$ is the electric field in the metal of the LSP mode, which we find by numerical means (using the commercial finite element software Comsol Multiphysics) by plane wave excitation with amplitude $E_0$ at resonance frequency $\omega_{\textrm{lsp}}$. 
The fact that probability of LSP excitation $P(\omega,\textbf{r})$ should decrease for frequencies that deviate from the resonance LSP frequency is accounted for by the function $N(\omega)$, which has the property that $N(\omega_{\textrm{lsp}})=1$ and decreases towards zero away from resonance. In this work, we extract $N(\omega)$ from calculation of the absorption spectrum of GSP-resonator arrays, but $N$ could also be approximated with a Lorentzian function whose center frequency and linewidth equals the resonance frequency and linewidth of the GSPR, respectively.
 
We can now setup a simple expression for the enhancement factor (EF) of PL intensity by the ratio of equations \ref{eq:PLfilm} and \ref{eq:PLns} integrated over the respective metal volumes, while taking into account all LSPs (denoted by integer $i$) found in the system, i.e.,
\begin{equation}
\textrm{EF}(\omega_\textrm{ex},\omega)=\frac{Q_{\textrm{ns}}}{Q_{\textrm{film}}}\left[1+\frac{1}{\eta_0}\sum_i\beta^{(i)}\eta_{\textrm{lsp}}^{(i)}S^{(i)}N^{(i)}\right],
\label{eq:EF}
\end{equation}
where $Q_j(\omega_\textrm{ex})=\int_\textrm{metal}U(\omega_{ex},\mathbf{r})\mathrm{d}V$ is the power absorbed in the metal (subscript $j=\textrm{ns}$ or $\textrm{film}$), and $S$ is defined as 
\begin{equation}
S(\omega_\textrm{ex},\omega_{\textrm{lsp}})=\frac{1}{E_0^2Q_\textrm{ns}}\int_{\textrm{ns}}U(\omega_\textrm{ex},\mathbf{r})|\mathbf{E}(\omega_{\textrm{lsp}},\mathbf{r})|^2\mathrm{d}V,
\label{eq:S}
\end{equation} 
and characterizes the spatial overlap between the absorption distribution at excitation frequency and the mode intensity at the LSP frequency. The different field distributions are exemplified in insets of Figure \ref{fig:Figure4}, hereby visually underlining that (for a given LSP mode) the spatial overlap $S$ depends on the excitation polarization. Note that the factor in front of the square bracket in equation \ref{eq:EF} accounts for the increased absorption at excitation frequency due to an increased amount of metal in GSP-resonator arrays relative to the metal film or due to any resonance effects at the excitation frequency, while the first term inside the square brackets represents PL from direct radiative decay of excited electron-hole pairs. The sum inside the square brackets, on the other hand, describes the plasmonic contribution to PL, with the integer $i$ representing all different LSPs that can be excited. For example, in considering detection of the total PL from GSP-resonator arrays we must take into account the two orthogonal LSPs whose contribution to the PL spectrum depends on the excitation polarization (i.e., $S$ depends on excitation polarization via $U$). In the different case of an analyzer in front of the detector (see, e.g., Figure \ref{fig:Figure3}e), we calculate the PL response by properly weighting the contribution from the two LSP modes, with modes of orthogonal polarization with respect to the analyzer axis being completely suppressed. It is interesting to note that the plasmonic part in equation \ref{eq:EF} resembles recently derived formulas\cite{Chatdanai2014} that have been used to accurately account for experimental observations. As a final comment, it ought to be mentioned that the direct applicability of equation \ref{eq:EF} rests on the knowledge of the product $\beta^{(i)}\eta_{\textrm{lsp}}^{(i)}$ in which $\eta_{\textrm{lsp}}^{(i)}$, as previously stated, can be estimated from the associated optical cross sections. No such simple estimation of $\beta^{(i)}$ exists (to the best of our knowledge), but it can be used as a fit parameter to experimental PL spectra. Alternatively, as adopted in this work, we can rely on studies of PL from spheres\cite{Dulkeith2004} and nanorods\cite{Mustafa2012}, which conclude that the PL quantum yield related to plasmonic resonances is to a large degree independent of size and shape due to the counter-acting processes of decreasing probability of electron-hole pair relaxation via LSP excitation and increasing radiative decay of the LSP when the size of the nanostructure increases. As such, we can to a first approximation assume $\beta^{(i)}\eta_{\textrm{lsp}}^{(i)}$ constant, with a value of $6\cdot10^{-11}$ used throughout this work.

Having developed a simple expression for the plasmon-assisted PL enhancement (equation \ref{eq:EF}), we apply the formalism to the nanodisk and brick arrays studied in this work (Figures \ref{fig:Figure2}b,d and \ref{fig:Figure3}c,e).
Note that the dimensions of the nanoparticles used in calculations were chosen to demonstrate the plasmon-induced influence on PL when the LSP is scanned through the wavelength range $\sim$600--900\,nm. As such, the dimensions of the simulated GSP-resonator arrays were adjusted wrt. the fabricated arrays in order to match resonance wavelength. In fact, since the shapes of fabricated nanodisks and nanobricks were not ideal (Figures \ref{fig:Figure1}d and \ref{fig:Figure3}a), their dimensions are not well defined. Additionally, the thickness and refractive index of the evaporated silica spacer layer are difficult to control\cite{Michael2012}. Overall, the modeled PL enhancement is in good agreement with experimental values and reflects the observed spectral tendencies.
For nanodisk arrays, it is clear that the model predicts a maximum PL enhancement for LSP wavelengths near $~680$\,nm (Figure \ref{fig:Figure2}b), with an associated maximum polarization contrast of $\alpha\sim 0.1$ (Figure \ref{fig:Figure2}d). This is in line with experiments, though calculations predict a faster decrease in PL intensity for LSP wavelengths below $\sim680$\,nm; a fact that we relate to the additional damping included in calculations, hereby leading to too strong contribution from interband transitions. Importantly, it should be stressed that in calculations the plasmonic contribution to the PL intensity lies solely in the parameter $S$ (equation \ref{eq:S}), meaning that large plasmon-enhanced PL can be reached by strong field enhancement at the excitation and LSP wavelengths together with a significant spatial overlap of the corresponding fields. Finally, let us consider calculations of PL from nanobrick arrays. In agreement with experimental results, the total PL enhancement features two peaks when the two orthogonal LSPs are spectrally separated (Figure \ref{fig:Figure3}c), while detection of PL with analyzer along either of the two nanobrick axes confirms the possibility of strongly polarized PL (Figure \ref{fig:Figure3}e). Neglecting the smallest array of nanobricks due to a strongly damped short-axis LSP, the remaining two configurations (Figure \ref{fig:Figure3}e; red and green curves) show average polarization contrast of $\alpha\sim0.5$ at the long and short LSP wavelength, respectively.

In the above presented work, the engineering of metal PL has been demonstrated within the spectral range 600-900\,nm, which is not an arbitrary chosen wavelength interval, but ultimately defined by the band gap of gold, hereby determining the availability of electrons and holes to recombine at a particular wavelength. Furthermore, the strong plasmon-induced modulation of PL relies upon gold being a relative good conductor in the considered wavelength range. As such, the engineering of metal PL in a different spectral range requires the usage of another plasmonic metal with a suitable band gap. For example, silver with a band gap at $\sim 400$\,nm is a potential option for transferring PL engineering to the blue and green spectral range\cite{Alqudami20007}.  
     
\section{Conclusion}	
In summary, we have demonstrated the flexibility of the GSP-resonator configuration for engineering the spectral and polarization content of gold PL by defining, in a single electron-beam lithography step, the dimensions of arrayed gold nanoparticles atop a subwavelength-thin dielectric spacer on a optically thick gold layer. The configuration features LSP modes that are excited by relaxation of excited charge carriers in gold and resulting in PL enhancement (at the geometrically defined LSP wavelength) via the LSP radiative decay. Broad wavelength tunability of PL enhancement was demonstrated, for which limiting factors were identified to be interband damping of LSP mode (short wavelengths) and diminished LSP excitation resulting from decreasing mode confinement in metal (long wavelengths). The PL was found to be either weakly or strongly polarized depending whether nanoparticles exhibited degenerate (nanodisk) or non-degenerate (nanobrick) orthogonal LSP modes. Additionally, we have shown how the relative contributions of the two orthogonal LSP modes are influenced by the polarization of the pump light.  In order to understand the underlying physics behind plasmon-enhanced PL, we have developed a simple model that takes into account the two relaxation paths of excited electron-hole pairs in metallic nanostructures, while relating local field enhancement effects at the excitation wavelength to enhanced PL emission at the LSP wavelength through the spatial overlap of absorption and LSP intensity profiles. The model faithfully reproduces all the PL features observed in experiment, including the dependence on excitation polarization and the strongly polarized nature of PL from nanobrick arrays with good quantitative agreement on PL enhancement values. We believe that the presented systematic study clearly demonstrates the potential of PL engineering by utilizing plasmonic resonances, with the developed theoretical model allowing one to properly choose system parameters, thereby providing guidelines for the wavelength and polarization engineering of PL. The results obtained might, in our opinion, also find implications in engineering of two-photon luminescence\cite{Beermann2008} and cathodeluminescence\cite{Abajo2010}.     	
\pagebreak

\pagebreak
\newpage

\section{Methods}
\paragraph{Fabrication} Samples were prepared on a diced silicon substrate by e-beam evaporation of 100\,nm gold (Au) and RF-sputtering of 20\,nm silicon dioxide (SiO$_2$) with intermediate 3\,nm layers of Titanium (Ti), added for adhesive purposes. The sample were then spincoated with PMMA 2A 950k to $\sim$100nm thickness and prebaked for 90\,s at 180\,C$^\circ$. Arrays of 300\,nm period disks or bricks were then defined by electron beam lithography (JSM-6400LV JEOL), developed in Methyl isobutyl(MIBK): Isopropyl alcohol(IPA) 1:3 (30s) with subsequent IPA stopper (30s) and blow-dried with nitrogen. Finally, 3\,nm Ti/ 50\,nm Au was deposited by thermal evaporation, and lift-off was done by leaving sample in acetone overnight ($\sim$ 10h). The sample was removed from the acetone bath under constant rinse of IPA and subsequently blow-dried with nitrogen. Disk and brick arrays were prepared under two separate fabrication cycles.

\paragraph{PL measurement}
PL spectra of the prepared GSP-resonator arrays were obtained on a PL spectroscopy setup constructed on a IX73 microscope (Olympus). The arrays were excited  by a linearly polarized, pulsed laser (LDH-P-FA-530L Pico Quant) at 532\,nm wavelength and 40\,MHz repetition rate, resulting in 1.2\,mW average power incident on sample. A half-waveplate were inserted in excitation path for rotating excitation polarization. Excitation and back collection of PL  were done through a x100 IR objective(NA 0.95), defocused to ensure broad excitation of the center portion of the array. The defocused spot was continuously reproduced by defocusing on the plane gold film to achieve the same spot image observed on CCD (wrt. image markers). PL was separated from excitation by a dichroic mirror and subsequent LP filter (band edge 550\,nm) and fiber coupled to a grating spectrometer(QE65000 Ocean Optics). For measurements of fluorescence polarization, an analyzer was fitted infront of the fiber coupler.

\paragraph{Scattering measurement}
Scattering spectra were obtained using a dark-field spectroscopy setup on a BX51 microscope (Olympus). The arrays were illuminated by a halogen lamp at highly oblique angles $\sim$70$^\circ$, while scattered light, was collected in the range 0-64$^\circ$ by a x100 (NA 0.9) objective. The collection of scattered light was restricted to a 32\,\textmu m spot (centered on array) by a pinhole in the objective image plane and fiber coupled to a grating spectrometer (QE65000 Ocean Optics). For measurements of polarization of scattered light, an analyzer was inserted infront of the pinhole. Scattering spectra were obtained relative to a reflection measurement of a silver mirror.

\paragraph{Simulations}
All modeling results are performed using the commercially available finite-element software Comsol Multiphysics, version 4.3b. In the calculations, we only model a single unit cell of the two-dimensional GSP-resonator arrays, applying periodic boundary conditions on the vertical sides of the cell. The lower boundary of the simulation domain, representing the truncation of the optically thick gold substrate, behaves as a perfect electric conductor, while the air domain above the GSP-resonator array is truncated using a port boundary. The same port boundary is also used for excitation of plane waves that propagate normal to the gold surface. Throughout this paper, we use a unit cell period of 300nm, a dielectric spacer thickness of 20nm, and nanostructures of height 50nm. The spacer is assumed to be silicon dioxide with a constant permittivity of 2.1, while the permittivity of gold is described by interpolated experimental values\cite{Johnson1972}. The imaginary part of the gold permittivity is increased by four times as a way of modeling the increased electron relaxation rate due to surface scattering and grained boundaries, and the presence of 3-nm-thin Ti adhesion layers\cite{Chen2010,Pors2013_2}.


\begin{acknowledgement}
We acknowledge financial support for this work from the Danish Council for Independent Research (the FNU project, contract no. 12-124690) and the European Research Council, Grant 341054 (PLAQNAP).
\end{acknowledgement}


\begin{suppinfo}
Comparative PL spectra of disk arrays for excitation co- and cross- polarized to either horizontal or vertical analyzer (S1); scattering and PL spectral comparison for analyzer probing polarization along short or long axis of brick arrays (S2); PL spectra of brick arrays for excitation polarized along short or long axis (S3).
\end{suppinfo}

%\bibliography{acs-Manuscript-references}

\begin{mcitethebibliography}{34}
\providecommand*\natexlab[1]{#1}
\providecommand*\mciteSetBstSublistMode[1]{}
\providecommand*\mciteSetBstMaxWidthForm[2]{}
\providecommand*\mciteBstWouldAddEndPuncttrue
  {\def\EndOfBibitem{\unskip.}}
\providecommand*\mciteBstWouldAddEndPunctfalse
  {\let\EndOfBibitem\relax}
\providecommand*\mciteSetBstMidEndSepPunct[3]{}
\providecommand*\mciteSetBstSublistLabelBeginEnd[3]{}
\providecommand*\EndOfBibitem{}
\mciteSetBstSublistMode{f}
\mciteSetBstMaxWidthForm{subitem}{(\alph{mcitesubitemcount})}
\mciteSetBstSublistLabelBeginEnd
  {\mcitemaxwidthsubitemform\space}
  {\relax}
  {\relax}

\bibitem[Mooradian(1969)]{Mooradin1969}
Mooradian,~A. Photoluminescence of Metals. \emph{Phys. Rev. Lett.}
  \textbf{1969}, \emph{22}, 185--187\relax
\mciteBstWouldAddEndPuncttrue
\mciteSetBstMidEndSepPunct{\mcitedefaultmidpunct}
{\mcitedefaultendpunct}{\mcitedefaultseppunct}\relax
\EndOfBibitem
\bibitem[Apell et~al.(1988)Apell, Monreal, and Lundqvist]{Apell1988}
Apell,~P.; Monreal,~R.; Lundqvist,~S. Photoluminescence of Noble Metals.
  \emph{Phys. Scripta} \textbf{1988}, \emph{38}, 174--179\relax
\mciteBstWouldAddEndPuncttrue
\mciteSetBstMidEndSepPunct{\mcitedefaultmidpunct}
{\mcitedefaultendpunct}{\mcitedefaultseppunct}\relax
\EndOfBibitem
\bibitem[Boyd et~al.(1986)Boyd, Yu, and Shen]{SurfaceRoughness}
Boyd,~G.~T.; Yu,~Z.~H.; Shen,~Y.~R. Photoinduced Luminescence from the Noble
  Metals and its Enhancement on Roughened Surfaces. \emph{Phys. Rev. B}
  \textbf{1986}, \emph{33}, 7923--7936\relax
\mciteBstWouldAddEndPuncttrue
\mciteSetBstMidEndSepPunct{\mcitedefaultmidpunct}
{\mcitedefaultendpunct}{\mcitedefaultseppunct}\relax
\EndOfBibitem
\bibitem[Lumdee et~al.(2014)Lumdee, Yun, and Kik]{Chatdanai2014}
Lumdee,~C.; Yun,~B.; Kik,~P.~G. Gap-Plasmon Enhanced Gold Nanoparticle
  Photoluminescence. \emph{ACS Photonics} \textbf{2014}, \emph{1},
  1224--1230\relax
\mciteBstWouldAddEndPuncttrue
\mciteSetBstMidEndSepPunct{\mcitedefaultmidpunct}
{\mcitedefaultendpunct}{\mcitedefaultseppunct}\relax
\EndOfBibitem
\bibitem[Lee et~al.(2004)Lee, Govorov, Dulka, and Kotov]{Jaebeom2004}
Lee,~J.; Govorov,~A.~O.; Dulka,~J.; Kotov,~N.~A. Bioconjugates of CdTe
  Nanowires and Au Nanoparticles:  Plasmon−Exciton Interactions,
  Luminescence Enhancement, and Collective Effects. \emph{Nano Lett}
  \textbf{2004}, \emph{4}, 2323--2330\relax
\mciteBstWouldAddEndPuncttrue
\mciteSetBstMidEndSepPunct{\mcitedefaultmidpunct}
{\mcitedefaultendpunct}{\mcitedefaultseppunct}\relax
\EndOfBibitem
\bibitem[Bouhelier et~al.(2003)Bouhelier, Beversluis, and
  Novotny]{Bouhelier2003}
Bouhelier,~A.; Beversluis,~M.~R.; Novotny,~L. Characterization of Nanoplasmonic
  Structures by Locally Excited Photoluminescence. \emph{Appl. Phys. Lett.}
  \textbf{2003}, \emph{83}, 5041--5043\relax
\mciteBstWouldAddEndPuncttrue
\mciteSetBstMidEndSepPunct{\mcitedefaultmidpunct}
{\mcitedefaultendpunct}{\mcitedefaultseppunct}\relax
\EndOfBibitem
\bibitem[Beermann et~al.(2008)Beermann, Novikov, S{\o}ndergaard, Boltasseva,
  and Bozhevolnyi]{Beermann2008}
Beermann,~J.; Novikov,~S.~M.; S{\o}ndergaard,~T.; Boltasseva,~A.;
  Bozhevolnyi,~S.~I. Two-photon mapping of localized field enhancements in thin
  nanostrip antennas. \emph{Opt. Express} \textbf{2008}, \emph{16},
  17302--17309\relax
\mciteBstWouldAddEndPuncttrue
\mciteSetBstMidEndSepPunct{\mcitedefaultmidpunct}
{\mcitedefaultendpunct}{\mcitedefaultseppunct}\relax
\EndOfBibitem
\bibitem[Bouhelier et~al.(2005)Bouhelier, Bachelot, Lerondel, Kostcheev, Royer,
  and Wiederrecht]{Bouhelier2005}
Bouhelier,~A.; Bachelot,~R.; Lerondel,~G.; Kostcheev,~S.; Royer,~P.;
  Wiederrecht,~G.~P. Surface Plasmon Characteristics of Tunable
  Photoluminescence in Single Gold Nanorods. \emph{Phys. Rev. Lett.}
  \textbf{2005}, \emph{95}, 267405\relax
\mciteBstWouldAddEndPuncttrue
\mciteSetBstMidEndSepPunct{\mcitedefaultmidpunct}
{\mcitedefaultendpunct}{\mcitedefaultseppunct}\relax
\EndOfBibitem
\bibitem[Beversluis et~al.(2003)Beversluis, Bouhelier, and Novotny]{Micael2003}
Beversluis,~M.; Bouhelier,~A.; Novotny,~L. Continuum generation from single
  gold nanostructures through near-field mediated intraband transitions.
  \emph{Phys. Rev. B} \textbf{2003}, \emph{68}, 115433\relax
\mciteBstWouldAddEndPuncttrue
\mciteSetBstMidEndSepPunct{\mcitedefaultmidpunct}
{\mcitedefaultendpunct}{\mcitedefaultseppunct}\relax
\EndOfBibitem
\bibitem[Mohamed et~al.(2000)Mohamed, Volkov, Link, and El-Sayed]{Mohamed2000}
Mohamed,~M.~B.; Volkov,~V.; Link,~S.; El-Sayed,~M.~A. The `Lightning' Gold
  Nanorods: Fluorescence Enhancement of over a Million Compared to the Gold
  Metal. \emph{Chem. Phys. Lett.} \textbf{2000}, \emph{317}, 517 -- 523\relax
\mciteBstWouldAddEndPuncttrue
\mciteSetBstMidEndSepPunct{\mcitedefaultmidpunct}
{\mcitedefaultendpunct}{\mcitedefaultseppunct}\relax
\EndOfBibitem
\bibitem[Dulkeith et~al.(2004)Dulkeith, Niedereichholz, Klar, Feldmann, von
  Plessen, Gittins, Mayya, and Caruso]{Dulkeith2004}
Dulkeith,~E.; Niedereichholz,~T.; Klar,~T.~A.; Feldmann,~J.; von Plessen,~G.;
  Gittins,~D.~I.; Mayya,~K.~S.; Caruso,~F. Plasmon Emission in Photoexcited
  Gold Nanoparticles. \emph{Phys. Rev. B} \textbf{2004}, \emph{70},
  205424\relax
\mciteBstWouldAddEndPuncttrue
\mciteSetBstMidEndSepPunct{\mcitedefaultmidpunct}
{\mcitedefaultendpunct}{\mcitedefaultseppunct}\relax
\EndOfBibitem
\bibitem[Shahbazyan(2013)]{Shahbazyan2013}
Shahbazyan,~T.~V. Theory of Plasmon-Enhanced Metal Photoluminescence.
  \emph{Nano Lett.} \textbf{2013}, \emph{13}, 194--198\relax
\mciteBstWouldAddEndPuncttrue
\mciteSetBstMidEndSepPunct{\mcitedefaultmidpunct}
{\mcitedefaultendpunct}{\mcitedefaultseppunct}\relax
\EndOfBibitem
\bibitem[Walsh and Negro(2013)Walsh, and Negro]{Walsh2013}
Walsh,~G.~F.; Negro,~L.~D. Engineering Plasmon-Enhanced Au Light Emission with
  Planar Arrays of Nanoparticles. \emph{Nano Lett.} \textbf{2013}, \emph{13},
  786--792\relax
\mciteBstWouldAddEndPuncttrue
\mciteSetBstMidEndSepPunct{\mcitedefaultmidpunct}
{\mcitedefaultendpunct}{\mcitedefaultseppunct}\relax
\EndOfBibitem
\bibitem[Steiner et~al.(2008)Steiner, Debus, Failla, and Meixner]{Mathias2008}
Steiner,~M.; Debus,~C.; Failla,~A.~V.; Meixner,~A.~J. Plasmon-Enhanced Emission
  in Gold Nanoparticle Aggregates. \emph{J.Phys.Chem.C} \textbf{2008},
  \emph{112}, 3103--3108\relax
\mciteBstWouldAddEndPuncttrue
\mciteSetBstMidEndSepPunct{\mcitedefaultmidpunct}
{\mcitedefaultendpunct}{\mcitedefaultseppunct}\relax
\EndOfBibitem
\bibitem[Fang et~al.(2012)Fang, Chang, Willingham, Swanglap, Dominquez-Medina,
  and Link]{Ying2012}
Fang,~Y.; Chang,~W.-S.; Willingham,~B.; Swanglap,~P.; Dominquez-Medina,~S.;
  Link,~S. Plasmon Emission Quantum Yield of Single Gold Nanorods as a Function
  of Aspect Ratio. \emph{ACS Nano} \textbf{2012}, \emph{6}, 7177--7184\relax
\mciteBstWouldAddEndPuncttrue
\mciteSetBstMidEndSepPunct{\mcitedefaultmidpunct}
{\mcitedefaultendpunct}{\mcitedefaultseppunct}\relax
\EndOfBibitem
\bibitem[Varnavski et~al.(2003)Varnavski, Mohamed, El-Sayed, and
  Goodson]{Oleg2003}
Varnavski,~O.~P.; Mohamed,~M.~B.; El-Sayed,~M.~A.; Goodson,~T. Relative
  Enhancement of Ultrafast Emission in Gold Nanorods. \emph{J. Phys. Chem. B}
  \textbf{2003}, \emph{107}, 3101--3104\relax
\mciteBstWouldAddEndPuncttrue
\mciteSetBstMidEndSepPunct{\mcitedefaultmidpunct}
{\mcitedefaultendpunct}{\mcitedefaultseppunct}\relax
\EndOfBibitem
\bibitem[S\"onnichsen et~al.(2002)S\"onnichsen, Franzl, Wilk, von Plessen,
  Feldmann, Wilson, and Mulvaney]{Sonnichsen2002}
S\"onnichsen,~C.; Franzl,~T.; Wilk,~T.; von Plessen,~G.; Feldmann,~J.;
  Wilson,~O.; Mulvaney,~P. Drastic Reduction of Plasmon Damping in Gold
  Nanorods. \emph{Phys. Rev. Lett.} \textbf{2002}, \emph{88}, 077402\relax
\mciteBstWouldAddEndPuncttrue
\mciteSetBstMidEndSepPunct{\mcitedefaultmidpunct}
{\mcitedefaultendpunct}{\mcitedefaultseppunct}\relax
\EndOfBibitem
\bibitem[Tcherniak et~al.(2011)Tcherniak, Dominguez-Medina, Chang, Swanglap,
  Slaughter, Landes, and Link]{Alexei2011}
Tcherniak,~A.; Dominguez-Medina,~S.; Chang,~W.-S.; Swanglap,~P.;
  Slaughter,~L.~S.; Landes,~C.~F.; Link,~S. One-Photon Plasmon Luminescence and
  Its Application to Correlation Spectroscopy as a Probe for Rotational and
  Translational Dynamics of Gold Nanorods. \emph{J. Phys. Chem. C}
  \textbf{2011}, \emph{115}, 15938--15049\relax
\mciteBstWouldAddEndPuncttrue
\mciteSetBstMidEndSepPunct{\mcitedefaultmidpunct}
{\mcitedefaultendpunct}{\mcitedefaultseppunct}\relax
\EndOfBibitem
\bibitem[Yorulmaz et~al.(2012)Yorulmaz, Khatua, Zijlstra, Gaiduk, and
  Orrit]{Mustafa2012}
Yorulmaz,~M.; Khatua,~S.; Zijlstra,~P.; Gaiduk,~A.; Orrit,~M. Luminescence
  Quantum Yield of Single Gold Nanorods. \emph{Nano lett.} \textbf{2012},
  \emph{12}, 4385--4391\relax
\mciteBstWouldAddEndPuncttrue
\mciteSetBstMidEndSepPunct{\mcitedefaultmidpunct}
{\mcitedefaultendpunct}{\mcitedefaultseppunct}\relax
\EndOfBibitem
\bibitem[Hu et~al.(2013)Hu, Akimov, Duan, Li, Liao, Tan, Wu, Chen, Fan, Bai,
  Lee, Yang, and Shen]{HailongHu2013}
Hu,~H.; Akimov,~Y.~A.; Duan,~H.; Li,~X.; Liao,~M.; Tan,~R. L.~S.; Wu,~L.;
  Chen,~H.; Fan,~H.; Bai,~P.; Lee,~P.~S.; Yang,~J. K.~W.; Shen,~Z.~X.
  Photoluminescence via Gap Plasmons between Single Silver Nanowires and a Thin
  Gold Film. \emph{Nanoscale} \textbf{2013}, \emph{5}, 12086--12091\relax
\mciteBstWouldAddEndPuncttrue
\mciteSetBstMidEndSepPunct{\mcitedefaultmidpunct}
{\mcitedefaultendpunct}{\mcitedefaultseppunct}\relax
\EndOfBibitem
\bibitem[Nielsen et~al.(2012)Nielsen, Pors, Albrektsen, and
  Bozhevolnyi]{Michael2012}
Nielsen,~M.~G.; Pors,~A.; Albrektsen,~O.; Bozhevolnyi,~S.~I. Efficient
  Absorption of Visible Radiation by Gap Plasmon Resonators. \emph{Opt.
  Express} \textbf{2012}, \emph{20}, 13311--13319\relax
\mciteBstWouldAddEndPuncttrue
\mciteSetBstMidEndSepPunct{\mcitedefaultmidpunct}
{\mcitedefaultendpunct}{\mcitedefaultseppunct}\relax
\EndOfBibitem
\bibitem[Pors and Bozhevolnyi(2013)Pors, and Bozhevolnyi]{pors2013_1}
Pors,~A.; Bozhevolnyi,~S.~I. Efficient and broadband quarter-wave plates by
  gap-plasmon resonators. \emph{Opt. Express} \textbf{2013}, \emph{21},
  2942--2952\relax
\mciteBstWouldAddEndPuncttrue
\mciteSetBstMidEndSepPunct{\mcitedefaultmidpunct}
{\mcitedefaultendpunct}{\mcitedefaultseppunct}\relax
\EndOfBibitem
\bibitem[Roberts et~al.(2014)Roberts, Pors, Albrektsen, and
  Bozhevolnyi]{roberts2014}
Roberts,~A.~S.; Pors,~A.; Albrektsen,~O.; Bozhevolnyi,~S.~I. Subwavelength
  Plasmonic Color Printing Protected for Ambient Use. \emph{Nano Lett.}
  \textbf{2014}, \emph{14}, 783--787\relax
\mciteBstWouldAddEndPuncttrue
\mciteSetBstMidEndSepPunct{\mcitedefaultmidpunct}
{\mcitedefaultendpunct}{\mcitedefaultseppunct}\relax
\EndOfBibitem
\bibitem[Hu et~al.(2012)Hu, Duan, Yang, and Shen]{HailongHu2012}
Hu,~H.; Duan,~H.; Yang,~J.~K.; Shen,~Z.~X. Plasmon-Modulated Photoluminescence
  of Individual Gold Nanostructures. \emph{ACS Nano} \textbf{2012}, \emph{6},
  10147--10155\relax
\mciteBstWouldAddEndPuncttrue
\mciteSetBstMidEndSepPunct{\mcitedefaultmidpunct}
{\mcitedefaultendpunct}{\mcitedefaultseppunct}\relax
\EndOfBibitem
\bibitem[Bozhevolnyi and S{\o}ndergaard(2007)Bozhevolnyi, and
  S{\o}ndergaard]{Bozhevolnyi2007}
Bozhevolnyi,~S.~I.; S{\o}ndergaard,~T. General Properties of Slow-Plasmon
  Resonant Nanostructures: Nano-Antennas and Resonators. \emph{Opt. Express}
  \textbf{2007}, \emph{15}, 10869--10877\relax
\mciteBstWouldAddEndPuncttrue
\mciteSetBstMidEndSepPunct{\mcitedefaultmidpunct}
{\mcitedefaultendpunct}{\mcitedefaultseppunct}\relax
\EndOfBibitem
\bibitem[Yuan et~al.(2007)Yuan, Chettiar, Cai, Kildishev, Boltasseva, Drachev,
  and Shalaev]{yuan2007}
Yuan,~H.-K.; Chettiar,~U.~K.; Cai,~W.; Kildishev,~A.~V.; Boltasseva,~A.;
  Drachev,~V.~P.; Shalaev,~V.~M. A negative permeability material at red light.
  \emph{Opt. Express} \textbf{2007}, \emph{15}, 1076--1083\relax
\mciteBstWouldAddEndPuncttrue
\mciteSetBstMidEndSepPunct{\mcitedefaultmidpunct}
{\mcitedefaultendpunct}{\mcitedefaultseppunct}\relax
\EndOfBibitem
\bibitem[Johnson and Christy(1972)Johnson, and Christy]{Johnson1972}
Johnson,~P.~B.; Christy,~R.~W. Optical Constants of the Noble Metals.
  \emph{Phys. Rev. B} \textbf{1972}, \emph{6}, 4370--4379\relax
\mciteBstWouldAddEndPuncttrue
\mciteSetBstMidEndSepPunct{\mcitedefaultmidpunct}
{\mcitedefaultendpunct}{\mcitedefaultseppunct}\relax
\EndOfBibitem
\bibitem[Pors et~al.(2010)Pors, Willatzen, Albrektsen, and
  Bozhevolnyi]{Pors2010}
Pors,~A.; Willatzen,~M.; Albrektsen,~O.; Bozhevolnyi,~S.~I. From plasmonic
  nanoantennas to split-ring resonators: tuning scattering strength. \emph{J.
  Opt. Soc. Am. B} \textbf{2010}, \emph{27}, 1680--1687\relax
\mciteBstWouldAddEndPuncttrue
\mciteSetBstMidEndSepPunct{\mcitedefaultmidpunct}
{\mcitedefaultendpunct}{\mcitedefaultseppunct}\relax
\EndOfBibitem
\bibitem[Viarbitskaya et~al.(2013)Viarbitskaya, Teulle, Marty, Sharma, Girard,
  Arbouet, and Dujardin]{Viarbitskaya2013}
Viarbitskaya,~S.; Teulle,~A.; Marty,~R.; Sharma,~J.; Girard,~C.; Arbouet,~A.;
  Dujardin,~E. Tailoring and Imaging the Plasmonic Local Density of States in
  Crystalline Nanoprisms. \emph{Nat. Mater.} \textbf{2013}, \emph{12},
  426--432\relax
\mciteBstWouldAddEndPuncttrue
\mciteSetBstMidEndSepPunct{\mcitedefaultmidpunct}
{\mcitedefaultendpunct}{\mcitedefaultseppunct}\relax
\EndOfBibitem
\bibitem[Alqudami and Annapoorni(2007)Alqudami, and Annapoorni]{Alqudami20007}
Alqudami,~A.; Annapoorni,~S. Fluorescence From Metallic Silver and Iron
  Nanoparticles Prepared by Exploding Wire Technique. \emph{Plasmonics}
  \textbf{2007}, \emph{2}, 5--13\relax
\mciteBstWouldAddEndPuncttrue
\mciteSetBstMidEndSepPunct{\mcitedefaultmidpunct}
{\mcitedefaultendpunct}{\mcitedefaultseppunct}\relax
\EndOfBibitem
\bibitem[Garc{\'\i}a~de Abajo(2010)]{Abajo2010}
Garc{\'\i}a~de Abajo,~F.~J. Optical excitations in electron microscopy.
  \emph{Rev. Mod. Phys.} \textbf{2010}, \emph{82}, 209--275\relax
\mciteBstWouldAddEndPuncttrue
\mciteSetBstMidEndSepPunct{\mcitedefaultmidpunct}
{\mcitedefaultendpunct}{\mcitedefaultseppunct}\relax
\EndOfBibitem
\bibitem[Chen et~al.(2010)Chen, Drachev, Borneman, Kildishev, and
  Shalaev]{Chen2010}
Chen,~K.-P.; Drachev,~V.~P.; Borneman,~J.~D.; Kildishev,~A.~V.; Shalaev,~V.~M.
  Drude Relaxation Rate in Grained Gold Nanoantennas. \emph{Nano Lett.}
  \textbf{2010}, \emph{10}, 916--922\relax
\mciteBstWouldAddEndPuncttrue
\mciteSetBstMidEndSepPunct{\mcitedefaultmidpunct}
{\mcitedefaultendpunct}{\mcitedefaultseppunct}\relax
\EndOfBibitem
\bibitem[Pors et~al.(2013)Pors, Nielsen, and Bozhevolnyi]{Pors2013_2}
Pors,~A.; Nielsen,~M.~G.; Bozhevolnyi,~S.~I. Broadband plasmonic half-wave
  plates in reflection. \emph{Opt. Lett.} \textbf{2013}, \emph{38},
  513--515\relax
\mciteBstWouldAddEndPuncttrue
\mciteSetBstMidEndSepPunct{\mcitedefaultmidpunct}
{\mcitedefaultendpunct}{\mcitedefaultseppunct}\relax
\EndOfBibitem
\end{mcitethebibliography}

\providecommand*\mcitethebibliography{\thebibliography}
\csname @ifundefined\endcsname{endmcitethebibliography}
  {\let\endmcitethebibliography\endthebibliography}{}

\newpage

%Pictures

\end{document}